\documentclass
[
    sigconf = false,     
    review = false,      
    screen = true,      
    anonymous = false,   
]
{acmart}

\AtBeginDocument{%
  \providecommand\BibTeX{{%
    \normalfont B\kern-0.5em{\scshape i\kern-0.25em b}\kern-0.8em\TeX}}}

\setcopyright{rightsretained}
\acmConference[Preprint]{}{arXiv}{2021}
\acmDOI{}
\acmISBN{}
\copyrightyear{2021} 

\usepackage{soul,longtable,lscape,subfigure} %
\usepackage{xcolor,hyperref}
\usepackage{algorithm}
\usepackage{algpseudocode}
\usepackage{tikz,pgfplots}

\begin{document}

\title{Addressing the Multiplicity of Solutions in Optical Lens Design as a Niching Evolutionary Algorithms Computational Challenge}

\author{Anna V. Kononova}
\orcid{0002-4138-7024}
\affiliation{
  \institution{LIACS, Leiden University}
  \country{The Netherlands}}
\email{a.kononova@liacs.leidenuniv.nl}

\author{Ofer M. Shir}
\affiliation{%
  \institution{Computer Science Department, Tel-Hai College, and Migal Institute}
  \city{Upper Galilee}
  \country{Israel}}
\email{ofersh@telhai.ac.il}

\author{Teus Tukker}
\affiliation{
  \institution{ASML}
  \city{Veldhoven}
  \country{The Netherlands}}
\email{teus.tukker@asml.com}

\author{Pierluigi Frisco}
\affiliation{
  \institution{ASML}
  \city{Veldhoven}
  \country{The Netherlands}}

\author{Shutong Zeng}
\affiliation{
  \institution{LIACS, Leiden University}
  \country{The Netherlands}}

\author{Thomas B{\"a}ck}
\affiliation{
  \institution{LIACS, Leiden University}
  \country{The Netherlands}}

\renewcommand{\shortauthors}{Kononova et al.}

\begin{abstract}
Optimal Lens Design constitutes a fundamental, long-standing real-world optimization challenge. Potentially large number of optima, rich variety of critical points, as well as solid understanding of certain optimal designs per simple problem instances, provide altogether the motivation to address it as a niching challenge. This study applies established Niching-CMA-ES heuristic to tackle this design problem (6-dimensional \textit{Cooke triplet}) in a simulation-based fashion. 
The outcome of employing Niching-CMA-ES `\textit{out-of-the-box}' proves successful, and yet it performs best when assisted by a local searcher which accurately drives the search into optima. The obtained search-points are corroborated based upon concrete knowledge of this problem-instance, accompanied by gradient and Hessian calculations for validation. We extensively report on this computational campaign, which overall resulted in (i) the location of 19 out of 21 known minima within a single run, (ii) the discovery of 540 new optima. These are new minima similar in shape to  21 theoretical solutions, but some of them have better merit function value (unknown heretofore), (iii) the identification of numerous infeasibility pockets throughout the domain (also unknown heretofore). 
We conclude that niching mechanism is well-suited to address this problem domain, and hypothesize on the apparent multidimensional structures formed by the attained new solutions. 
\end{abstract}



\maketitle

\sloppy{

\section{Introduction}
The domain of Optical Lens Design \cite{HGross} offers a variety of highly complex optimization problem instances: the solution spaces are very rugged, and the extensive mathematical techniques are of limited use in most of the cases. The conventional approach followed in designing lenses is to create a few designs using one's own experience, then let commercial packages (e.g., Code V \cite{CodeV} and OpicStudio \cite{OpticStudio}) to optimize these designs using gradient-based local search. Different techniques have been used to determine lens designs. For example, \cite{Vasiljevic:texbook} gives an overview of classical and Evolutionary Algorithms (EAs) for the optimization of optical systems, \cite{HoullierLepine2019} compares different global optimization algorithms on freeform optical designs,  \cite{HoschelLaksh2018} reviews the use of genetic algorithms for lens design, and different mathematical techniques are employed to locate the minima of specific lens designs in \cite{BaciortEO2004, GLorEO2009}.

Generally speaking, most of the approaches aim to locate the global optima, whereas only a few are designed to find local optima (see, e.g., \cite{GLorEO2009}). As mentioned earlier, because many factors (manufacturability, cost, etc,) are unknown a-priori, and have the potential to influence the choice of the final design, it makes sense to come up with methodologies searching for as many local optima as possible having some key performance indicator (KPI) values below/above given thresholds\footnote{The so-called \emph{Second Toyota Paradox}, which is often reviewed in management studies, promotes the consideration of multiple candidate solutions during the car production process \cite{Cristiano}:
\begin{quote}
\emph{"Delaying decisions, communicating ambiguously, and pursuing
an excessive number of prototypes, can produce better cars faster
and cheaper."}
\end{quote}
}.

In this study we propose a novel approach for the challenge of locating as many local minima as possible in the domain of optical lens system designs (i.e., optical systems). 
Our proposed approach is based on the application of a niching algorithm that relies on a variant of the Covariance Matrix Adaptation Evolution Strategy (CMA-ES) \cite{Shir-NACO08},  combined with local search. 

Lens designs are subject to input and output constraints: the type of light the optical system has to deal with, the magnification/reduction factor the optical system must achieve, the focal length the optical system must have, the occupied volume and manufacturability tolerances, and the maximum cost of the system. The KPIs of a lens design can be many and they are related to aberrations, i.e., imperfections in image formation. These aberrations are computed with a ray tracing algorithm which models how the optical design affects the light passing through it. 

\section{Problem formulation}
\subsection{Optical perspective}
Optical lenses are well-known physical objects that focus or diverge light. The path of rays of light hitting the lens in its different parts differs, as illustrated in Figure~\ref{fig:SpotSize} which shows a simple lens design with one lens and paths taken by two rays. The central ray (shown as a dashed black line) of an on-axis object point passes through the middle of the lens and hits the focal plane in the center. Another, off-center, ray (shown with a red line) is deflected by the lens and hits the focal plane center with a $x$ and $y$ displacement. The area on the focal plane reached by all the considered rays is called the spot. It is the size of this spot that characterises a lens (design). 
\begin{figure}
  \includegraphics[width=0.9\columnwidth,trim={7mm 0mm 6mm 4mm},clip]{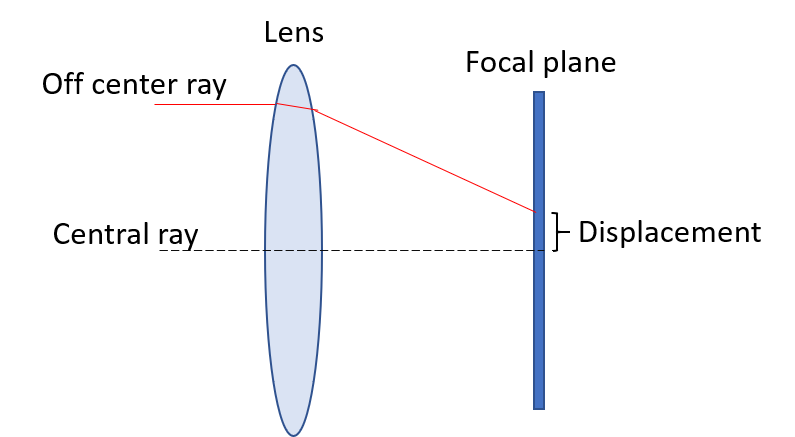}
  \caption{Depiction of a simple lens design where the central ray hits the center of the focal plane and an off-center ray hits the focal plane with a displacement from the desired center position.}\label{fig:SpotSize}
\end{figure}

\subsubsection{Statement of the problem}\label{sect:optics-statement}
A lens collects light emitted or reflected by an object and focuses it in the image plane. In the case of an ideal lens an exact scaled image of the object will appear. However, the image of a real lens will be deformed and blurred to some extent due to geometrical aberrations. Diffraction effects will be ignored in this paper as the geometrical errors are dominating.   

\begin{figure}
  \includegraphics[width=0.68\columnwidth,trim={7mm 5mm 3mm 6mm},clip]{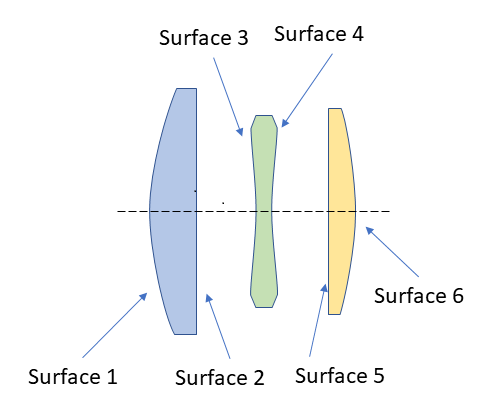}
  \caption{Example of Cooke triplet with the indication of the surface numbers.}\label{fig:coockeTriplet}
\end{figure}
A \textit{Cooke triplet} is an optical system that consists of 3 lens elements placed consecutively. The design problem associated with the Cooke Triplet consists of adjusting six lens curvatures $\mathbf{c} = (c_1,\ldots,c_6)$ (i.e., two surfaces per lens, see Figure~\ref{fig:coockeTriplet}) in order to obtain the optical system with the best imaging properties (an optical system is \textit{imaging} if it focuses part of the light emitted by points on a plane in the object space into a point on a plane in the image space). This lens system is well known and used in optical design theory as it has just enough free parameters to correct all primary aberrations. In this work the lens and air gap thicknesses are not varied as these have only small contributions to the lens performance. Next to this the glass material is not varied to reduce the complexity of the design problem as was done in reference \cite{GLorEO2009}.

The imaging properties of a system can be quantified
as the root mean square (RMS) of the spot size: 
\begin{equation}
    S_1(\mathbf{c}) = \sqrt{\frac{1}{n} \sum_{i=1}^{n}{(\Delta_i(\mathbf{c}))^2}} \rightarrow \min
\end{equation}\label{eq:S}
%
where $\mathbf{c}$ the vector of lens curvatures of the given optical system and $\Delta_i(\mathbf{c}) = \Delta x_i(\mathbf{c}) + \Delta y_i(\mathbf{c})$ are the displacements in $x$- and $y$-coordinates 

Typically, to compute the spot size of a given system, a limited number of rays originating from a small number of object heights\footnote{A line can be drawn through center of the lens (see the dotted line in Figure~\ref{fig:coockeTriplet}) extents from the object to the image. The object height is then the distance of a point to that axis.} in this system needs to be simulated and traced. Such approach has been taken here: tracing 126 rays originating from 3 different object heights. Then, the distance of the point of incidence on the image plane to the geometrical center of the spot was calculated for each of these rays and added to the merit function. 

\subsubsection{Optics simulators} Tracing of rays of light passing through an optical system can be done with several optical simulators, with subtle differences. Commonly used commercial simulators are CODE V \cite{CodeV} and OpticStudio \cite{OpticStudio}. OpticStudio has been chosen here because of the availability of code to interact with MATLAB \cite{MATLAB}. A small difference between the two simulators, however, needed to be addressed. OpticStudio does not allow setting optimisers in front of the optical stop as was used in  \cite{GLorEO2009}. 
A workaround is that after the new curvature parameters are set and before the KPI is computed a MATLAB script sets the object distance such that the lens system has a magnification of value of $-1$.


\subsubsection{Typical approach to solve lens design problem taken by optical designers.}  One of the approaches often taken by experienced optical designers is to start with a known design and improve upon it by manually adding or removing lens elements or even groups of elements with a known optical `function'. After each alteration the system parameters are optimized with a merit function containing the key KPI's, typically with a local optimizer such as the Damped Least Square method. Time need for a design and the final lens performance largely depend on the \textit{experience of the designer}. At the same time, since the employed optimisers are operating locally there is a high probability that good designs are missed. 

Therefore, strictly speaking, \textit{no directly competitive computational approaches exist} in the field of optics for the methodology proposed in this paper. 

\subsubsection{Infeasible lens designs}
Clearly, some lens designs can result in rays falling outside the focal plane or reflecting internally. For such designs, ray tracing simulators return a predefined large numerical value.

\subsection{Computational perspective}\label{sect:perspective}
The lens design problem described in the previous section constitutes a (reasonably) low-dimensional continuous problem set within a simple hypercube domain whose objective function\footnote{Referred to as KPI in the previous sections} $f(\mathbf{c})$ is based on a number of outputs obtained from an optics simulator\footnote{The objective function used here is the same as in \cite{GLorEO2009}.} $S$: 
%
\begin{eqnarray}
    f(\mathbf{c}) & = & w_1 S_1(\boldsymbol{c}) \\ \nonumber
    & & + w_2 (S_{\text{EFFL}}(\boldsymbol{c}) - \text{EFFL}_\text{target})^2 \\
    & & + w_3 (S_{\text{PMAG}}(\boldsymbol{c}) - \text{PMAG}_\text{target})^2 \rightarrow \min \;, \nonumber 
\end{eqnarray}\label{eq:objf}
\noindent where $\mathbf{c}=(c_1,\ldots,c_6)\in [-0.25,0.25]^6$ are lens curvatures, $S_1(\boldsymbol{c})$ is the spot size merit function component defined in Section~\ref{sect:optics-statement}, $S_{\text{EFFL}}(\boldsymbol{c})$ is the effective focal length component, and $S_{\text{PMAG}}(\boldsymbol{c})$ is the paraxial magnification, all three being computed by the simulator $S$. The target values of the latter two are set as $\text{EFFL}_\text{target} = 30 [\text{mm}]$ and $\text{PMAG}_\text{target} = -1$, respectively, and the weights are chosen as $w_i = 1$. \textit{Minimisation problem specified in eq.~(2) is solved in this paper.} 


\subsubsection{Analysis of known optima}
The Cooke triplet lens design problem has been intensively studied in the past. To date, the most detailed analysis of the problem \cite{GLorEO2009} yields curvature values which represent  $21$ locally optimal designs. 
This analysis is a mixture of numerical simulation and physics-based theoretical reasoning about the properties of a simplified objective function. 
The authors suggest a methodology of constructing triplet minima from known duplet minima (and, potentially for higher orders, incrementally) and build a classification of locally optimal designs according to their morphology. 

Their results, however, do not exclude the possibility of existence of other local optima. \textit{The approach taken in this paper is to study how close a specialised evolutionary algorithm can get to the known optima in a purely numerical fashion, without utilizing any optics knowledge.} 
\subsubsection{Discrepancies between simulators} It should be mentioned that results in \cite{GLorEO2009} have been obtained using the Code V simulator. Meanwhile, in this study the OpticsStudio simulator has been employed. To assure the compatibility of the two simulators, a study has been carried out where $21$ known optima have been evaluated in both simulators - results coincide up to the $5^\text{th}$ decimal digit.

\subsubsection{Nature of Critical Points}\label{sec:classical_methods}
As described above, previous investigation of this optimization domain, and particularly of this triplet problem-instance, concluded that the underlying search landscape possesses a rich variety of critical points \cite{GLorEO2009}. Since we study the application of randomized search heuristics in a black-box fashion, an assessment of the attained solution-points is much needed in order to validate their nature as optima. To this end, we consider finite-differences calculations of the gradient vector and the Hessian matrix \cite{NumericalRecipes}, relying on the short evaluation times of the simulator's objective function calls. Explicitly, a critical point of an objective function $f$ is where its gradient vector vanishes, i.e., $\left( \nabla f \right)_i:= \frac{\partial f}{\partial x_i} =0~\forall i$. The Hessian of $f$, $\left(\mathcal{H}_f\right)_{ij} := \frac{\partial^2 f}{\partial x_i \partial x_j}$ may provide the diagnosis of the critical points by examination of its eigenvalues (which are guaranteed to be real due to its symmetry): a positive (negative) spectrum indicates a minimum (maximum), whereas a spectrum with mixed signs indicates a saddle point; the existence of a zero eigenvalue suggests flatness and may require higher-order derivatives for assessment. 

In practice, gradients and Hessian spectra of the 21 known theoretical solutions were approximated using finite-differences' calculations, and by utilizing a variety of infinitesimal perturbation values (delta steps) to validate the approximation ($\delta x := 10^{-4},~10^{-5},\ldots,10^{-12}$). The obtained gradients indeed reflect the critical point nature of these solutions. Furthermore, we analyzed the attained Hessian spectra. Evidently, the majority of the eigenvalues vanish –- indicating some flatness, but formally suggesting lack of conclusive statements. All spectra are dominated by a single direction -- which has been corroborated as a meaningful design by the human experts -- exhibiting high condition numbers at these basins of attraction. Notably, two optima appear to be saddle points (i.e., possess negative eigenvalues). This odd observation may be explained by the discrepancies between the two simulators, suggesting that these critical-points' locations (verified on Code-V) undergo a shift on OpticStudio.

\begin{figure}
  \centering
  \includegraphics[width=0.73\columnwidth,angle=270,trim={0mm 3mm 1mm 10mm},clip]{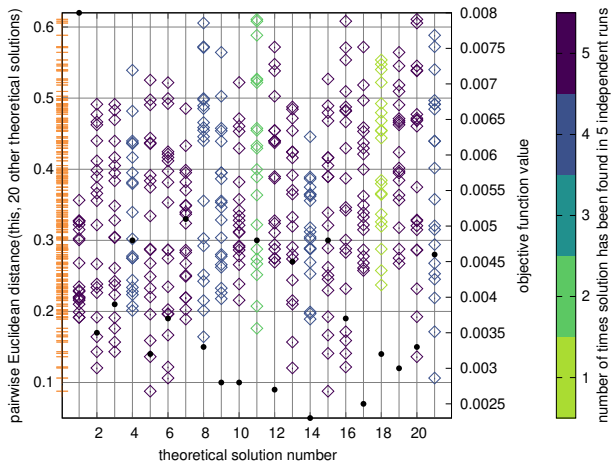}
  \caption{Distribution of pairwise Euclidean distances among the known 21 theoretical solutions (6-dimensional real-valued decision vectors that are associated with 6 lens curvatures), as identified in \cite{GLorEO2009}, having the solutions' numbering preserved. 
  Orange dashes indicate the distribution of all $\frac{21\times20}{2}$ pairwise distances among 21 theoretical solutions, i.e. they are projections of all rhombi onto the left vertical axis. The objective function value of each theoretical solution is depicted as a black circle, with values marked on the right vertical axis. 
  Solution number 14 is the global minimum. The rhombi symbols are colored according to the number of times a theoretical solution has been precisely located in a series of 5 independent runs of the final configuration of the proposed method shown in Algorithm ~\ref{alg:approach}.}
  \label{fig:distances}
\end{figure}
\subsubsection{Known Optima Locations}\label{sect:known_locations} 
As amply demonstrated in the field of computational intelligence \cite{Kononova2015,Davarynejad2014,Caraffini2019,Kononova2020CEC,Kononova2020PPSN}, location of optima within the domain can influence the performance of heuristic optimisation algorithms. Therefore, here we investigate how 21 known optima \cite{GLorEO2009} are distributed within the problem domain and look for additional insights into the structure of the problem. 

As described above, the domain corresponding to the Cooke Triplet problem is a hypercube in six dimensions, of real-valued parameters. Instead of searching for efficient visualisations of two- or three-dimensional projections and complex multidimensional reasoning, we opt for a study of the distribution of pairwise distances between know optima\footnote{Euclidean distance is chosen here as in this relatively low dimensionality, it still allows good distinction between distances \cite{Aggarwal2001}. Switching to Manhattan distance yields very similar results.} -- see Figure~\ref{fig:distances} and its caption containing a detailed explanation of the figure. Careful inspection of this figure allows the following conclusions:
\begin{itemize}
    \item pairwise distances between theoretical solutions vary in the range $[0.08,0.62]$; most pairwise distances lie in the range $[0.2,0.5]$; 
    \item the global optimum (solution number 14) is located overall further away from other good solutions; at the same time, all other solutions lie at similar distances to the global solution; similar distribution can be observed for the worst performing optimum (solution 1);
    \item other better solutions (17, 12, 9, 10) are also located further away from other remaining solutions;
    \item no direct correlation between the distribution of distances for similar performing solutions can be observed (e.g., solutions 5 vs 18 or 6 vs 16);
    \item in a series of 5 independent runs of the final configuration of the proposed method discussed in Section~\ref{sect:method}:
    \begin{itemize}
        \item[$\star$] all solutions have been found \textit{at least once}; in total, 18 to 19 out of 21 solutions have been found in every single run; 
        \item[$\star$] the least found solution (solution 18, found 1 time out of 5) is located further away from all other solutions;
        \item[$\star$] the second least found solution (solution 11, found 2 times out of 5) is not drastically different from other solutions.
    \end{itemize}
\end{itemize}

\section{Methodology and setup}\label{sect:method}
In this section we elaborate on the computational approach taken, the preliminary planning, and the experimentation setup. We first present the background of niching methods and specify the concrete methodology that we employ in this research.

\subsection{Niching CMA-ES}
Standard Evolutionary Algorithms (EAs) tend to lose their population diversity and converge into a single solution \cite{ShirHandbookNACO}. \emph{Niching methods} constitute the extension of EAs to finding multiple optima in relevant search-landscapes within one population \cite{ShirHandbookNACO}. They address this issue by maintaining the diversity of certain properties within the population, and thereby aim at obtaining parallel convergence into multiple basins of attraction in the landscape within a single run. Research on niching methods started in Genetic Algorithms \cite{MahfoudPHD}, followed by work in Evolution Strategies \cite{Shir-NACO08,Shir-SA_ECJ}, and broadened to the entire field of nature-inspired heuristics yielding altogether a sheer volume of potent techniques (see \cite{NichingSurvey2017} for a recent review).

Here, given the optimization problem at hand, we choose to employ the established, radius-based Niching CMA-ES technique \cite{Shir-NACO08} using a $\left(1,\lambda\right)$ kernel. The targeted number of niches is denoted by $q$. In short, following the evaluation of the population, niches are spatially constructed around peak individuals, in a \emph{greedy} manner, based upon the prescribed niche radius $\rho$. Resources are uniformly partitioned per niche, thus each peak individual is sampled $\lambda$ times in the following generation\footnote{In socio-biological terms, the peak individual is associated with an {\bf alpha-male}, which wins the local competition and gets all the sexual resources of its ecological niche. The algorithm as a whole can be thus considered as a competition between alpha-males, each of which is fighting for one of the available $q$ ``computational resources'', after winning its local competition at the ``ecological optimum'' site.}. 

\subsection{Parameter Settings}\label{sect:params}
We employ the niching within a $\left(1,\lambda\right)$-CMA-ES algorithm, and specify in what follows its parameters' settings by adhering to the notation in \cite{Shir-NACO08}. We target $q=20$ niches per each run, yet allocate further $p=5$ dynamic peaks (i.e., 25 D-sets are formed). Regarding population sizing, we follow the recommendation and set $\lambda=10$ -- yielding altogether 250 individuals that undergo evaluation in each iteration. Importantly, setting the niche radius value may have a critical impact on the behavior of a niching routine that features a fixed radius, often referred to as the \textit{niche radius problem} \cite{Shir-SA_ECJ}. Here, rather than approximating the search volume and partitioning it among the niches \cite{Shir-SA_ECJ}, we are in a position to capitalize on the known theoretical solutions. 
We assume that their spatial distribution is indicative of the general distribution of solution points within the feasible space, and set accordingly the niche radius to half of the mean pairwise distances (see Figure \ref{fig:distances}): $\rho=0.18$. Otherwise, the cycle of non-peak reset is set to $\kappa=20$ iterations, and the initial global step-size is set to $\sigma_0=0.05$. Solutions generated outside the domain are corrected by placing them on the boundary \cite{Kononova2020PPSN}.

\subsection{Local Search Utilization}\label{sect:LS}
The so-called Dampened Least Squares (DLS) \cite{Vasiljevic:texbook,Guo19} (also known as the Levenberg-Marquardt algorithm) constitutes a modification of the well-known Newton-Raphson method \cite{NumericalRecipes}. It is traditionally employed in lens design, and it became a built-in option in the OpticsStudio simulator. This local search method on its own, being dependent upon `good' initial points, is unlikely to locate all the optima of the problem. We thus utilised it for validation purposes, as explained in Section \ref{sec:classical_methods}. At the same time, preliminary experiments suggested that minor improvements are often achieved when this local search method is applied to the solution-points attained by the niching algorithm, regardless of the niching configuration and its parameters' setting. We explain these observed improvements by the landscape's rich variety of critical points, including saddle points, which seemingly render the convergence attempts by the CMA-ES challenging. We therefore devised a \textit{hybrid} approach, as outlined in Algorithm~\ref{alg:approach}.
\begin{algorithm}[!ht]
\begin{algorithmic}[1]
	\State{$i\leftarrow1$}
	\State{$a \leftarrow \emptyset$}\Comment{initialise solution archive}
    \State{$o \leftarrow \emptyset$}\Comment{initialise optima archive}
    \State{$d \leftarrow 10$}\Comment{set archive search depth}
	\State{$p[i] \leftarrow$ initialise niching $\left(1,\lambda\right)$-CMA-ES}\Comment{see Section~\ref{sect:params}}
	\While{fitness evaluation budget permits}
        \State{$p[i+1]\leftarrow$ $i^{th}$ generation of niching $\left(1,\lambda\right)$-CMA-ES} \Comment{\cite{Shir-NACO08}}
		\State{$a[i] \leftarrow p[i]$} \Comment{update solution archive}
		\State{$i\leftarrow i+1$}
	\EndWhile
	\For{$j = i-d \rightarrow i-1$}\Comment{take last $d$ generations}
	    \While{$a[j]$ is not empty}
	        \State{$s \leftarrow$ fetch next element of $a[j]$}
            \State{$o\leftarrow o \cup DLS(s$)} \Comment{via  OpticsStudio\footnotemark}
        \EndWhile
	\EndFor
	\State \Return filter$(o)$\Comment{remove duplicates, return optima archive}
\caption{The Proposed Niching-$\left(1,\lambda\right)$-CMA-ES + DLS Hybrid Approach}\label{alg:approach}
\end{algorithmic}
\end{algorithm}
\footnotetext{Apply DLS implementation from OpticStudio until the internal termination criteria is not hit: no significant improvement in the value of the objective function.}

\subsection{Setup and Experimental Planning}
Our niching implementation follows the publicly available source code \cite{NichingCMAESsourcecode}.
The niching algorithm was run using the configuration from Section~\ref{sect:params} for up to 25000 objective function evaluations, with the variation operator being adjusted within the predefined 6-dimensional boundaries. 
An approximate duration of a single objective function call is within the range of 2 \texttt{sec}.
All the experiments were run using \texttt{MATLAB} and executed on Windows Intel(R) Core(R)i5 CPU 8350  @ 1.90GHz with 4 processing units. 
The final Niching-$\left(1,\lambda\right)$-CMA-ES configuration has been run 5 times, meanwhile overall, we conducted 60 runs with various settings.



\section{Results and Analyses}
Our proposed approach located altogether 540 optima within this domain. Furthermore, during the reported experimentation, \textit{all the evaluated} candidate solution-points (i.e., lens designs) visited by the aforementioned approach have been recorded. Upon filtering out duplicate points, 185200 unique feasible candidate solution-points were altogether located, versus 245880 infeasible candidate solution-points. Next, we elaborate on these observations and offer insights into the landscape. 

\subsection{Distribution of Known and Novel Optima}
The aforementioned experimentation has consistently indicated the existence of new locally optimal solutions. Overall, \textit{540 new solutions} have been identified and further investigated for local optimality based on critical-points' analyses (Section \ref{sec:classical_methods}). In particular, their gradients were approximated, and were shown to reflect their critical-point nature. 
\begin{figure}
  \centering
  \includegraphics[width=0.73\columnwidth,angle=270,trim={0mm 2mm 1mm 8mm},clip]{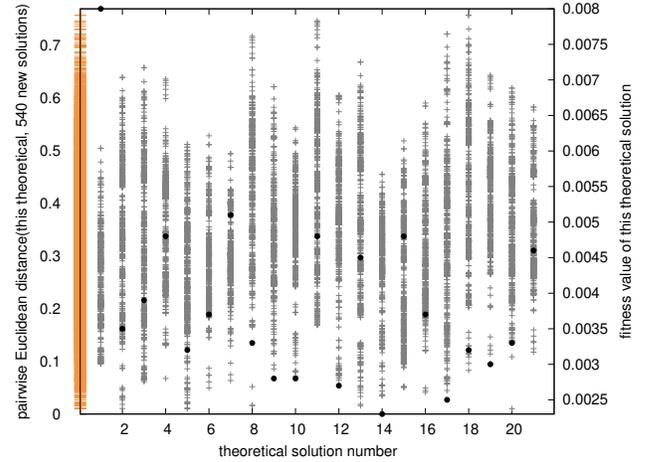}
  \caption{Distances between each theoretical and all new solutions, overlaid with fitness values of theoretical solutions shown in filled black circles with values on the right vertical axis. Moreover, distributions of distances for solutions  are shown with $540\times21$ dashes in orange.}
  \label{fig:distances_th2new_euc}
\end{figure}

\begin{figure*}
  \centering
  \includegraphics[width=1.5\columnwidth,angle=270,trim={3mm 2mm 1mm 8mm},clip]{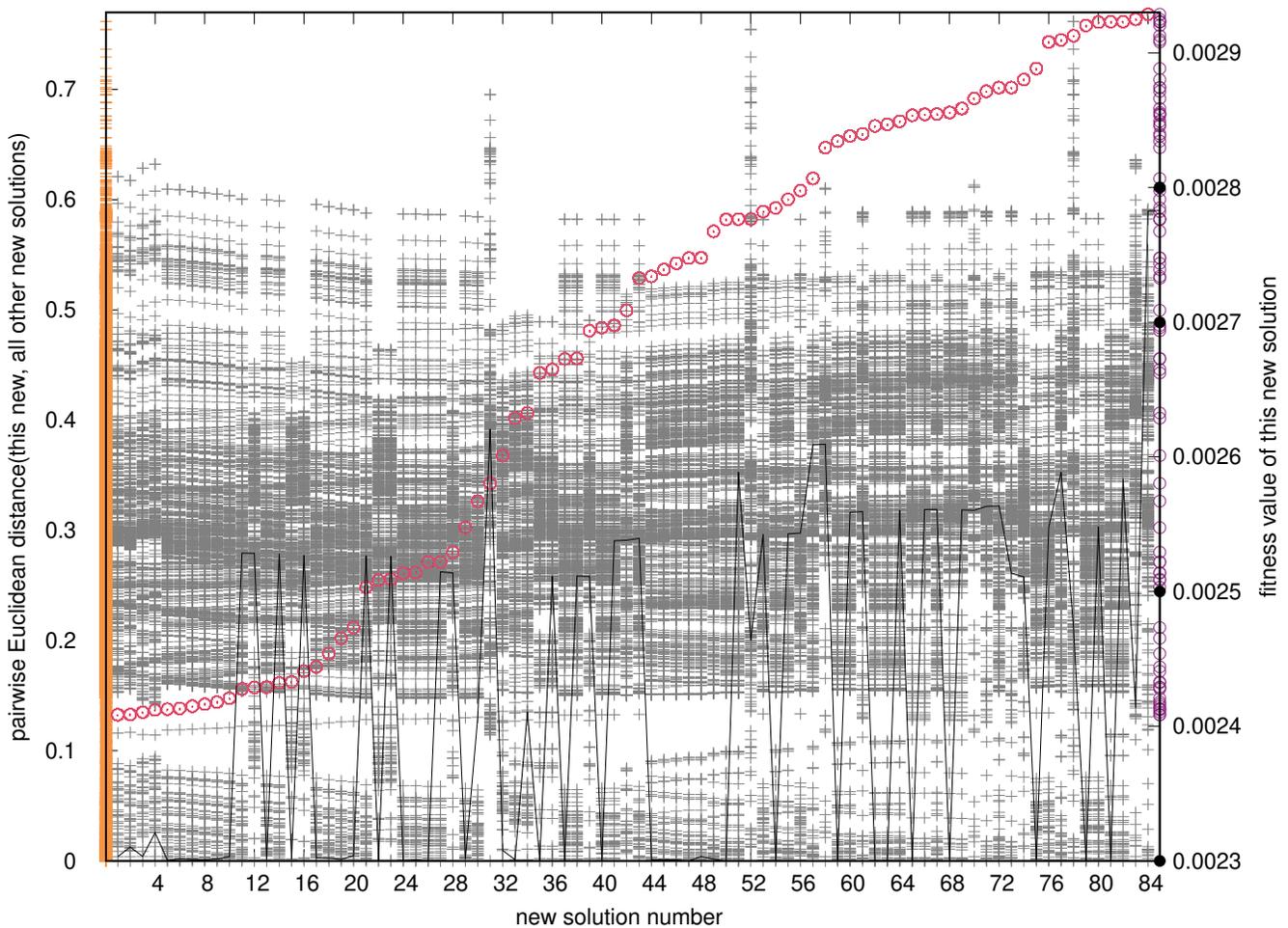}
  \vspace{-2mm}
  \caption{Distribution of pairwise Euclidean distances between new solutions (6 lens curvatures or 6-dimensional real-valued vectors) is shown as 539 grey pluses for each of the 84 best performing solutions out of total 540 new solutions found. Overlaid on top are (as empty pink circles, with values on the right vertical axis) fitness values of these 84 solutions and (as black line, with values on the left vertical axis) distances to the next best solution. Left and right vertical axes also show, respectively, distributions of distances for all 84 new solutions (with $539\times84$ dashes in orange; these are projections of all grey pluses onto this axis) and fitness values of these solutions (with $84$ empty circles in violet; these are projections of pink circles onto this axis). Moreover, for comparison, fitness values of $4$ best performing theoretical solutions are shown with filled black circles on the right vertical axis. New solutions are sorted according to their fitness.}
  \label{fig:distances_new}
\end{figure*}

Finally, similarly to what has been reported in Section~\ref{sect:known_locations}, we study the distribution of pairwise Euclidean distances among 84 (out of 540, selected best in terms of attained objective function values) newly identified solutions (Figure~\ref{fig:distances_new}) and among the same 84 new solutions and 21 known theoretical solutions discussed in Section~\ref{sect:perspective} (Figure~\ref{fig:distances_th2new_euc}). Based on these two Figures, we are in a position to articulate the following conclusions:
\begin{itemize}
    \item (see Figure~\ref{fig:distances_th2new_euc}) pairwise distances among new solutions densely vary in the range $[0,0.77]$; (see Figure~\ref{fig:distances_new}) pairwise distances among all new and known theoretical solutions vary in the similar range $[0,0.76]$; i.e. in general new solutions lie at similar distances to each other as to the theoretical solutions;
    \item (see Figure~\ref{fig:distances_th2new_euc}) some of the new solutions lie close to theoretical solutions 2, 5, 8, 13, 14 (global minimum) and 17; majority of new solutions lie far away from known theoretical solutions; new solutions lie furthest away from theoretical solution 11; 
    \item (see Figure~\ref{fig:distances_new}) no new global optimum is found, but 20 new solutions outperform the second-best theoretical solution (``first runner-up''); 
    \item (see Figure~\ref{fig:distances_new}) objective function values of the 540 new solutions smoothly fill the range $[0.0024,0.00293]$, except for some `jumps' (e.g., between the $20^\text{th}$ and the $21^\text{st}$ best new solutions).
    \item (see Figure~\ref{fig:distances_new}) with a few exceptions, similarly-performing new solutions lie at similar distances to all new solutions (e.g. solutions 5--11 or 44--49);
    \item (see Figure~\ref{fig:distances_new}) relatively few new solutions are `clustered' together in terms of distance -- no new solution exhibits predominance of zero distances;
    \item (see Figure~\ref{fig:distances_new}) almost no new better performing solutions lie at the distance of about 0.12 from each other; other similar \textit{`exclusion zones'} can be observed at larger distances; 
    \item (see Figure~\ref{fig:distances_new}) many new solutions lie at the distance of about 0.3 from each other; other similar \textit{`density zones'} can be observed at larger distances; 
    \item (see Figure~\ref{fig:distances_new}) 3 out of the 84 solutions shown here lie further away from some new solutions (solutions 31, 52 and 78); these solutions do not differ in terms of objective function values when compared to the rest;
    \item (see Figure~\ref{fig:distances_new}) when sorted by their objective function values, each subsequent new solution on average equally either lies at distance of nearly 0.0 or at some distance in the range $[0.25,0.4]$;
\end{itemize}


\begin{figure}
    \centering
    \begin{tikzpicture}[scale=.71]
        \node (A) at (0, 0.2) {$245880$ sampled solutions};
        \node (B) at (0, -0.9) {$109328$ infeasible inside domain};
        \node (C) at (6.2, -0.9) {$\approx 6\times10^{9}$ p/w distances };
        \node (D) at (0, -2) {${500}$ subsampled infeasible $|^{100}_{i=1}$};
        \node (E1) at (-1.5, -3.5) {$1^{st}$min$|^{100}_{i=1}$};
        \node (E2) at (0.5, -3.5) {...};
        \node (E3) at (2.5, -3.5) {$10^{th}$min$|^{100}_{i=1}$};
        \node (F) at (6.2, -2) {$\approx 1.2\times10^5$ p/w distances $|^{100}_{i=1}$};
        \node (G) at (6.2, -3.5) {$5000$ subsamples $|^{100}_{i=1}$};
        \node (H) at (6.2, -4.7) {Shapiro-Wilk p-values $|^{100}_{i=1}$};
        \node (J) at (1.5, -4.7) {Figure~\ref{fig:dist_infeas}};
        \draw[->,orange] (A) edge (B) (B) edge (C) (B) edge (D) (F) edge (E1) (F) edge (E2) (F) edge (E3) (D) edge (F) (G) edge (H) (F) edge (G) (H) edge (J) (E1) edge (J) (E2) edge (J) (E3) edge (J);
    \end{tikzpicture}
    \vspace{-5mm}
    \caption{A schematic explanation for samples of 10 lowest distances and Shapiro-Wilk p-values calculation that are shown in Figure~\ref{fig:dist_infeas}; `p/w' stands for `pairwise' and $|^{100}_{i=1}$ denotes the number of repetitions.}\label{fig:explain}
\end{figure}
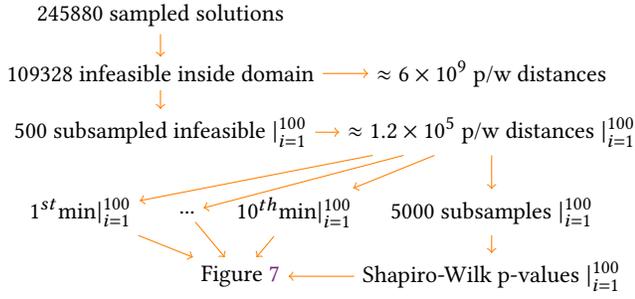

\subsection{Infeasibility Map of the Landscape}
Given the proportion of located infeasible points, we are in a position to state that this problem domain has \textit{inner infeasibility pockets}, wherein the objective function cannot be computed -- the simulator exhausts its computational budget and returns a predefined high value. Such infeasibility pockets are \textit{uniformly} distributed within the domain: 109328 out of the 245880 sampled solutions were located strictly within the domain boundaries. Their existence, and particularly their large volume\footnote{following filtering and removal of duplicate solutions, 57\% of the evaluated designs turned out to be infeasible.}, constitute a new observation in the field of optical design, where \textit{such pockets have not been noted heretofore}. 

How large are these infeasibility pockets? 
Clearly, their true size can only be \textit{estimated} based on a limited albeit large sample of 109328 unique evaluated solutions. 
It is not practical to study pairwise distances between all these solutions. 
Instead, we systematically (see Figure~\ref{fig:explain} for explanation) down-sample 100 times 500 infeasible solutions\footnote{i.e., on average $\frac{500}{109328}\approx 0.46$\% of the  infeasible solutions fall in each such sample} from 109328 points in the inner part of domain, compute 10 minimum values\footnote{the lowest, the second lowest, up to the 10$^\text{th}$ lowest} of pairwise Euclidean distances within each such sample\footnote{i.e., on average $\frac{10\times2}{500\times499}\approx 0.008$\% of computed distances are examined per sample}, plot them and report $p$-values from the normality test on distances within one sample (based on a subsample of 5000 distance values\footnote{i.e., the test per sample is run for $\frac{5000\times2}{500\times499}\approx$0.4\% of computed distances}). 
From the results presented in Figure~ \ref{fig:dist_infeas}, we are able to clearly state the following:
\begin{itemize}
    \item Infeasibility pockets predominantly consist of individual solutions. 
    This can be seen from only a small number of points with zero/low $1^\text{st}$ minimal distance and steadily increasing values of $2^\text{nd}$, ... , $10^\text{th}$ minimal distances. However, the possibility of existence of infeasibility pockets of nonzero measure is not excluded due to a very small sample size (in proportion) and the fact that the increase of subsequent minimal distances (see Figure~ \ref{fig:dist_infeas}) takes place very slowly.
    \item According to the Shapiro-Wilk normality tests with $\alpha=0.05$, subsampled pairwise distances among infeasible points are normally distributed in 62\% samples of size $5000$\footnote{see Figure~\ref{fig:dist_infeas}, 100\%-38\%.}.
    \item The values shown in Figure~\ref{fig:dist_infeas} are rather \textit{volatile}, i.e., they vary over multiple runs of the aforementioned procedure. 
    This happens due to small sample sizes (in proportion) subsampled from all $0.5\times109328\times109327\approx6\times10^9$ distances among infeasible solutions, see Figure~\ref{fig:explain}.
\end{itemize}
\begin{figure}
  \centering
  \includegraphics[width=0.72\columnwidth,angle=270,trim={2mm 2mm 0mm 8mm},clip]{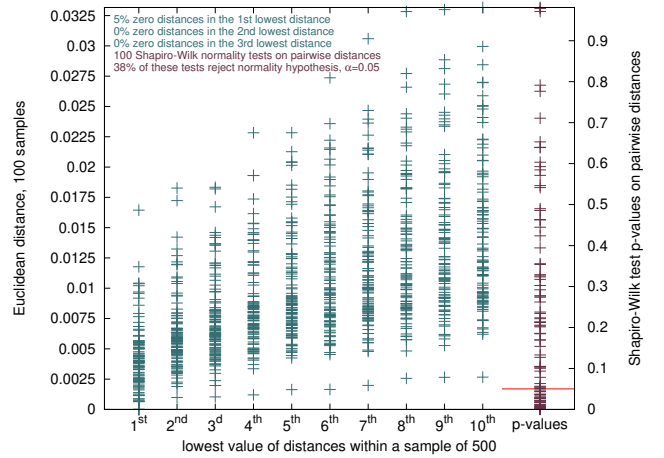}
  \vspace{-2mm}
  \caption{Distribution of lowest values of pairwise Euclidean distances and p-values from Shapiro-Wilk normality tests, $\alpha=0.05$. See Figure~\ref{fig:explain} for an explanation on the underlying calculation.}
  \label{fig:dist_infeas}
\end{figure}

Furthermore, we have constructed a binary `feasible/infeasible' \textit{classifier} for the standardised data based on the Support Vector Machine algorithm \cite{hastie2013elements}, featuring a polynomial kernel and 5-fold cross-validation -- which attains 85\% accuracy. 
However, \textit{details of this Machine Learning approach fall outside the scope of this paper}. 





\begin{figure}
\centering
\subfigure{\includegraphics[width=1.0\linewidth,trim={6mm 4mm 5mm 2mm},clip]{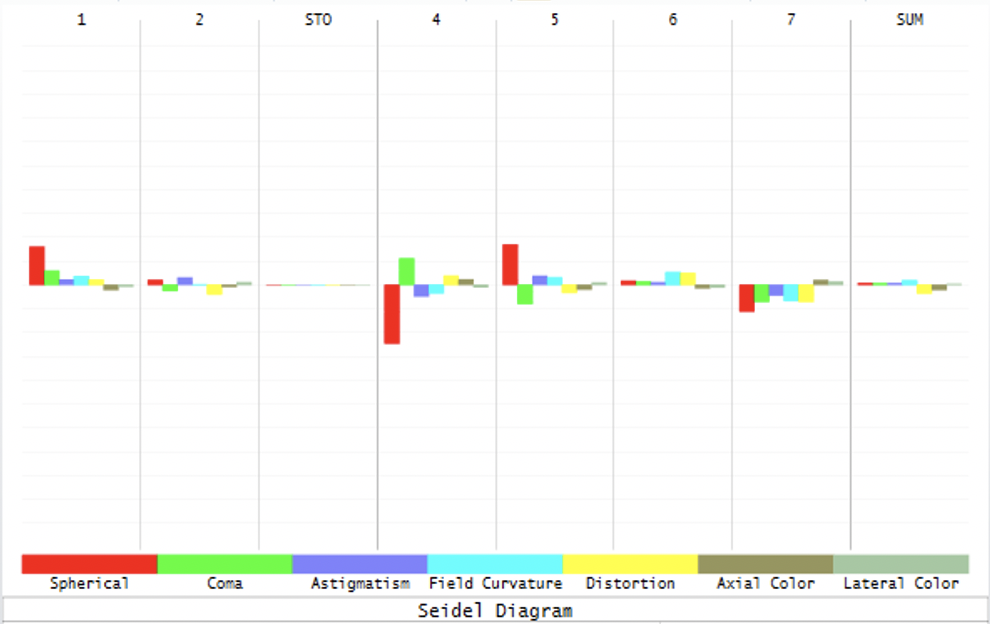}}\\
\vspace{-3mm}
\subfigure{\includegraphics[width=1.0\linewidth,trim={6mm 4mm 5mm 2mm},clip]{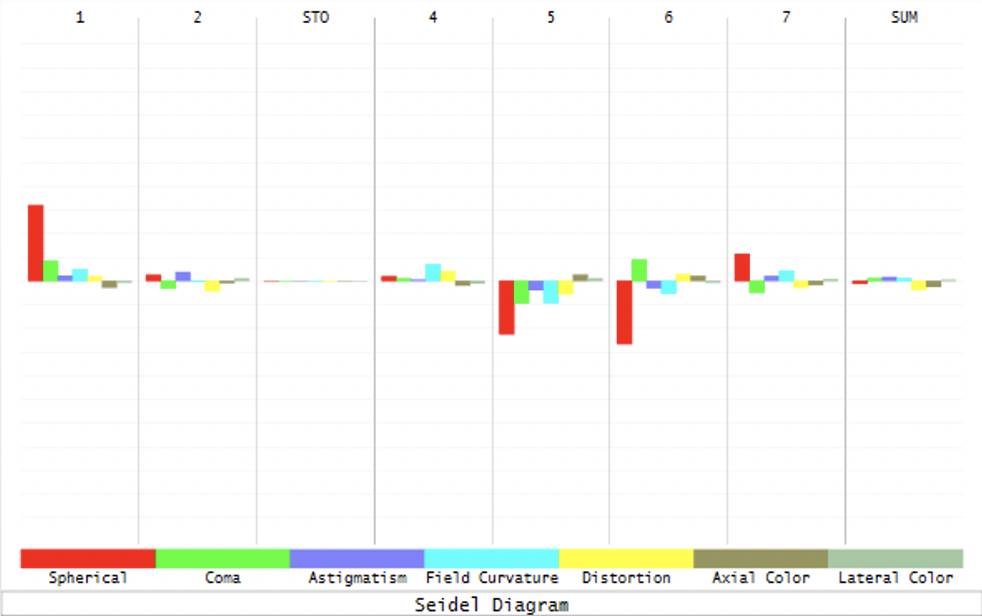}}
\caption{Seidel Diagram \cite{Seidel} from OpticStudio \cite{OpticStudio} indicating the main aberrations of two Cooke triplet systems: based on new solutions 1 and 84 - above and below, respectively (numbering of solutions is taken from Figure~\ref{fig:distances_new}). Columns labeled 1, 2 and 4-7 correspond to the six lens surfaces; for each such surface the spherical, coma, astigmatism, field curvature, distortion, axial and lateral color aberrations are shown with different colours. The height of each coloured bar shows the strength of the corresponding aberration. Maximum aberration scale is 0.5[mm], horizontal grid line spacing is 0.05[mm]. Rightmost column labeled SUM shows the total aberrations of the system. Obviously, the two solutions lead to a different behaviour of the system (different amount of aberrations in each lens). Both solutions are good as indicated by low sum aberrations.}\label{fig:optical}
\end{figure}

\subsection{Optical Perspective of the Novel Optima}
New solutions identified by the proposed approach have been investigated from the point of view of optics. At the first glance, all of them fall within the `morphological classes' specified in \cite{GLorEO2009} (i.e., they have the same sequence of the signs of the lens surface curvatures). However, within such class, their exact performance differs - Cooke triplets with curvatures specified by new solutions deliver numerically different aberrations \cite{Seidel} per lens, see Figure~\ref{fig:optical} for an example. Further detailed analysis of the new solutions from the optics point of view is outside the scope of this paper. 

Most of new solutions deliver not only low sum but also low individual (per lens) aberrations and, therefore, potentially can be used in industrial production and satisfy the aforementioned manufacturability constraints. Analysis of these individual aberrations has made it clear that within the approach proposed in this paper additional constraints, additional objective or penalty terms in the existing objective function \textit{can} be imposed to control such individual aberrations. This requires further study. 

\section{Conclusions}
In this work we studied a basic instance of the Optical Lens Design problem in light of the multiplicity of its solutions, and investigated the application of an existing niching approach to it. Our study unveiled a fascinating real-world application as a testbed for randomized search heuristics (and particularly black-box continuous solvers that operate on multimodal domains), and at the same time, provided insights on the search landscape of this application domain. Evidently, this low-dimensional optimization problem possesses an underlying rough landscape, which is likely to get worse if larger lens design problems are to be considered. Altogether, this problem-instance is significantly more complex than previously thought. Therefore, to address those challenges, specialized search heuristics are much needed to be devised.

The employed niching approach, which was utilized `out-of-the-box' using its defaults configuration, proved successful for this type of problem. During the current computational campaign, this approach typically obtained the majority of the known, theoretical solutions, and furthermore identified a few hundreds of new critical points. When compared to a commonly-utilized approach in this application domain, namely the saddle-point construction method, the niching approach is advantageous since it does not require a null (reference lens) element that does not change the merit function. This makes the method applicable to most optical design problems.

\balance
%


%
%
}
\bibliographystyle{ACM-Reference-Format}
\bibliography{biblio}


\end{document}